\documentclass[prd,superscriptaddress,amsmath,twocolumn,nofootinbib]{revtex4-2}

\pdfoutput=1
\usepackage{times}

\usepackage{graphicx}
\usepackage{xcolor}

\usepackage{array}
\usepackage{booktabs}
\usepackage{enumitem}
\usepackage{epsf,latexsym,bbm,euscript}

\usepackage{mathrsfs,mathtools,amsmath,amssymb}
\usepackage{tensor}

\makeatletter
\newcommand*{\defeq}{\mathrel{\rlap{%
			\raisebox{0.3ex}{$\m@th\cdot$}}%
		\raisebox{-0.3ex}{$\m@th\cdot$}}%
	=}
\newcommand*{\eqdef}{=\mathrel{\rlap{%
			\raisebox{0.3ex}{$\m@th\cdot$}}%
		\raisebox{-0.3ex}{$\m@th\cdot$}}%
}
\makeatother

\newcommand{\RNum}[1]{\uppercase\expandafter{\romannumeral #1\relax}}

\usepackage{scalerel}
\usepackage{tikz}
\usetikzlibrary{svg.path}
\definecolor{orcidlogocol}{HTML}{A6CE39}
\tikzset{
	orcidlogo/.pic={
		\fill[orcidlogocol] svg{M256,128c0,70.7-57.3,128-128,128C57.3,256,0,198.7,0,128C0,57.3,57.3,0,128,0C198.7,0,256,57.3,256,128z};
		\fill[white] svg{M86.3,186.2H70.9V79.1h15.4v48.4V186.2z}
		svg{M108.9,79.1h41.6c39.6,0,57,28.3,57,53.6c0,27.5-21.5,53.6-56.8,53.6h-41.8V79.1z M124.3,172.4h24.5c34.9,0,42.9-26.5,42.9-39.7c0-21.5-13.7-39.7-43.7-39.7h-23.7V172.4z}
		svg{M88.7,56.8c0,5.5-4.5,10.1-10.1,10.1c-5.6,0-10.1-4.6-10.1-10.1c0-5.6,4.5-10.1,10.1-10.1C84.2,46.7,88.7,51.3,88.7,56.8z};
	}
}
\newcommand\orcidlink[1]{\href{https://orcid.org/#1}{\mbox{\scalerel*{
				\begin{tikzpicture}[yscale=-1,transform shape]
					\pic{orcidlogo};
				\end{tikzpicture}
			}{X}}}}

\usepackage{url,hyperref}
\hypersetup{colorlinks,linkcolor={blue!55!black},citecolor={red!50!black},urlcolor={blue!45!black},breaklinks=true}

\allowdisplaybreaks
\begin{document}

\title{Elastic scattering of electron by a Yukawa potential in non-commutative spacetime}

\author{Abdellah Touati\orcidlink{0000-0003-4478-2529}}
\email{touati.abph@gmail.com}
\affiliation{Department of Physics, Faculty of Sciences and Applied Sciences, University of Bouira, Algeria}


\begin{abstract}
In this paper, we investigate the elastic scattering of an electron by a Yukawa potential within the framework of non-commutative (NC) geometry. We first derive the NC correction to the Yukawa potential at leading order in the NC parameter, resulting in a modified potential resembling a screened Kratzer potential. This potential reduces to the standard Kratzer form when considering the NC correction to the Coulomb potential. Subsequently, we calculate the NC correction to the electron scattering amplitude using the first-order Born approximation. We then analyze the effects of NC geometry on both the differential and total cross sections for elastic scattering. Our results indicate that non-commutativity enhances the differential cross section at small scattering angles and naturally gives rise to a Kratzer-like potential, reflecting the quantum nature of spacetime. Additionally, we establish a direct relationship between the system's energy level and the bound on the NC parameter. Specifically, for an ultra-relativistic incident electron scattering by a heavy molecule, we derive a new lower bound on $\Theta$ of the order of $10^{-28}\,\text{m}$.
\end{abstract}

\maketitle

\section{Introduction} \label{sec:introduction}

In particle and nuclear physics, the interaction of particles with nuclei has significant implications for both experimental and theoretical research \cite{scattering3,scattering4,scattering5}. These studies enhance our understanding of fundamental interactions, the composition of matter, and the properties of fundamental particles, including mass, spin, and electric charge, among others.

Potentials are regarded as accurate models for describing interactions between particles, atoms, and molecules. Notable examples include the Coulomb potential \cite{coulomb}, the screened Coulomb potential (Yukawa potential) \cite{Yukawa1}, and the Kratzer potential \cite{kratzer1}. Investigating the scattering of particles by such potentials provides a robust theoretical framework for modeling experimental observations \cite{scattering1,scattering3,scattering2,scattering4,scattering5}.

A key example of such interactions is the elastic scattering of electrons by atoms \cite{scattering1,yukawa3,yukawa4}, which has numerous applications, including electron microscopy. The scattering of electrons by different targets, such as atoms \cite{yukawa3}, ions \cite{scattering5}, molecules \cite{scattering6,scattering7}, and crystals \cite{crystals}, provides valuable insights into the physical properties of these systems. Several methods exist for solving scattering problems to determine cross sections, including the Plane-Wave method, which solves the differential equation for particle in the presence of a scattering potential \cite{cohen,cohen2}, and quantum field theory techniques based on Feynman diagrams \cite{greiner1,greiner2,greiner3}. One of the most widely used approaches is the Born approximation, which considers only the incident wave and the contributions from single scattering events with the potential \cite{sakurai}.

To date, the collision process incorporates three of the four fundamental interactions; electromagnetism, the weak, and the strong interaction, while excluding gravity, due to its weak coupling constant. To include gravity in such interactions, the gravity must be quantized. Various theories have emerged to describe gravity at the quantum scale, including string theory \cite{ST1,ST2}, loop quantum gravity \cite{LQG1,LQG2}, and supergravity \cite{SG1,SG2,SG3,SG4}. Another promising approach is non-commutative (NC) geometry \cite{noncommutative1}, which posits that quantizing spacetime leads to the quantization of gravity, a concept that originally emerged naturally in string theory \cite{seiberg1,berger}. This approach has attracted significant interest across various fields, including the standard model \cite{NCSM1,NCSM2,NCSM3,NCdecay1,NCcoll1,NCcoll2,NCcoll3,NCcoll4}, string theory \cite{berger}, the quantum Hall effect \cite{NCQHE1,NCQHE2}, and quantum field theory \cite{starproduct1,NCQFT1,NCfield1,NCfield2},...etc. Importantly, NC geometry plays a crucial role in quantum gravity by predicting a minimal length scale, which has significant implications for quantum systems.

In this paper, we extend the investigation of scattering processes within the framework of NC geometry, which allows us to explore the influence of quantum geometry on scattering phenomena. This provides an direct study of the quantum gravity effects on subatomic systems. In NC spacetime, the geometry is characterized by a modified commutation relation between the spacetime coordinates themselves \cite{noncommutative1}:
\begin{equation}
	[x^{\mu},x^{\nu}] = i\Theta^{\mu\nu},
\end{equation}
where $\Theta^{\mu\nu}$ is a real antisymmetric matrix. In NC geometry, the ordinary product between two arbitrary functions, $f(x)$ and $g(x)$, is replaced by the Moyal star product "$*$", defined as:
\begin{equation} \label{eq:star}
	(f * g)(x) = f(x) e^{\frac{i}{2} \Theta^{\mu\nu} \overleftarrow{\partial_{\mu}} \overrightarrow{\partial_{\nu}}} g(x).
\end{equation}
Using the Bopp's shift transformation and the first-order Born approximation, we investigate the total cross section for electron scattering by a Yukawa potential in NC spacetime, and study the relation between the energy level and the lower bound of the NC parameter.

In this study, we derive the NC correction to the Yukawa potential to leading order in the NC parameter. We then obtain the NC correction to the scattering amplitude for an electron in NC spacetime. Our results show that the NC Yukawa potential, as well as the NC Coulomb potential, correspond to the screened Kratzer potential and the standard Kratzer potential, respectively. We also analyze the impact of non-commutativity on the differential and total cross-sections for elastic scattering. Then we study and present a lower bound on the NC parameter.

This paper is organized as follows: In Sec.~\ref{sec:YPNCST}, we present the NC corrections to the Yukawa potential using Bopp's shift transformation. In Sec.~\ref{sec:NCCf}, we derive the NC corrections to the scattering amplitude for electrons in NC spacetime. In Sec.~\ref{sec:NCCS}, we examine the effect of NC geometry on the differential and total cross sections for elastic scattering. Finally, in Sec.~\ref{sec:concl}, we provide our concluding remarks.


\section{Yukawa potential in NC spacetime}\label{sec:YPNCST}

The Yukawa potential is a widely used potential in atomic collision studies, providing a good approximation for the Coulomb field of a nucleus screened by atomic electrons \cite{yukawa2,yukawa3}. It is also employed to describe the residual strong interaction between nucleons \cite{Yukawa1}. The Yukawa potential is given by the following expression:
\begin{equation}
	V(r)=\frac{V_0}{r}e^{-\alpha r},\label{eq:YP}
\end{equation}
where $V_0=-Z k_C e^2$, $k_C$ is the Coulomb constant, and $\alpha$ is the screening parameter \cite{scattering2}. To express this potential in the context of NC spacetime, we need to employ a coordinate transformation. For this purpose, we use Bopp's shift \cite{bopp's1}, which represents a shift in the coordinates and is defined as:
\begin{equation}
	\hat{x}^\mu=x^\mu-\frac{1}{2}\Theta^{\mu\nu}P_\nu,
\end{equation}
In this work, we consider spacetime non-commutativity with $\Theta^{i0}\neq0$, which is represented by the following NC matrix $\Theta^{\mu\nu}$:
\begin{equation}
	\Theta^{\mu\nu}=\left(\begin{matrix}
		0	& \Theta & 0 & 0 \\
		-\Theta	& 0 & 0 & 0 \\
		0	& 0 & 0 & 0 \\
		0	& 0 & 0 & 0
	\end{matrix}
	\right), \quad \mu,\nu=0,1,2,3\label{eq:NCM}\,.
\end{equation}
In spherical coordinates, this shift is expressed as:
\begin{equation}
	\hat{r}=r-\frac{1}{2}\Theta^{1 0}P_0,\label{eq:BSC2}
\end{equation}
where the component $P_0$ is defined by the correspondence principle as $P_0=i\hbar\partial_{0}$. In a conserved system, this quantity corresponds to the energy of the particle, i.e., $P_0=E/(\hbar c)$.
Next, we express the Yukawa potential in NC spacetime as:
\begin{equation}
	\hat{V}(\hat{r})=\frac{V_0}{\hat{r}}e^{-\alpha \hat{r}},\label{eq:YPNC1}
\end{equation}
By substituting the transformation \eqref{eq:BSC2} into Eq.~\eqref{eq:YPNC1} and expanding to the leading order in the NC parameter, we obtain the following expressions \cite{moumni1}:
\begin{subequations}
\begin{align}
	\hat{r}&=r-\frac{\Theta E}{2\hbar c},\\
	\frac{1}{\hat{r}}&\sim \frac{1}{r}+\frac{\Theta E}{2\hbar c r^2},
\end{align}
\end{subequations}
Inserting these relations into the potential expression \eqref{eq:YPNC1}, we derive the NC correction to the Yukawa potential as:
\begin{equation}
	\hat{V}(r)=e^{\frac{1}{2\hbar c}\alpha\Theta E}\left(\frac{V_0}{r}+\frac{V_0\Theta E}{2\hbar c r^2}\right)e^{-\alpha r},\label{eq:YPNC2}
\end{equation}
It is evident that in the commutative limit, i.e., $\Theta=0$, we recover the standard Yukawa potential \eqref{eq:YP}. The expression in Eq.~\eqref{eq:YPNC2} represents the NC Yukawa potential, which is structurally similar to the screened Kratzer potential form given in \cite{ikot1,Kratzer3}:
\begin{equation}
	\hat{V}(r)=\left(\frac{V_1}{r}+\frac{V_2}{r^2}\right)e^{-\alpha r},\label{eq:YPNC3}
\end{equation}
where $V_1=e^{\frac{1}{2}\alpha\Theta E}V_0$ and $V_2=\frac{V_0\Theta E}{2\hbar c} e^{\frac{1}{2\hbar c}\alpha\Theta E}$. 

\begin{figure*}[]
	\centering
	\includegraphics[width=0.25\textwidth]{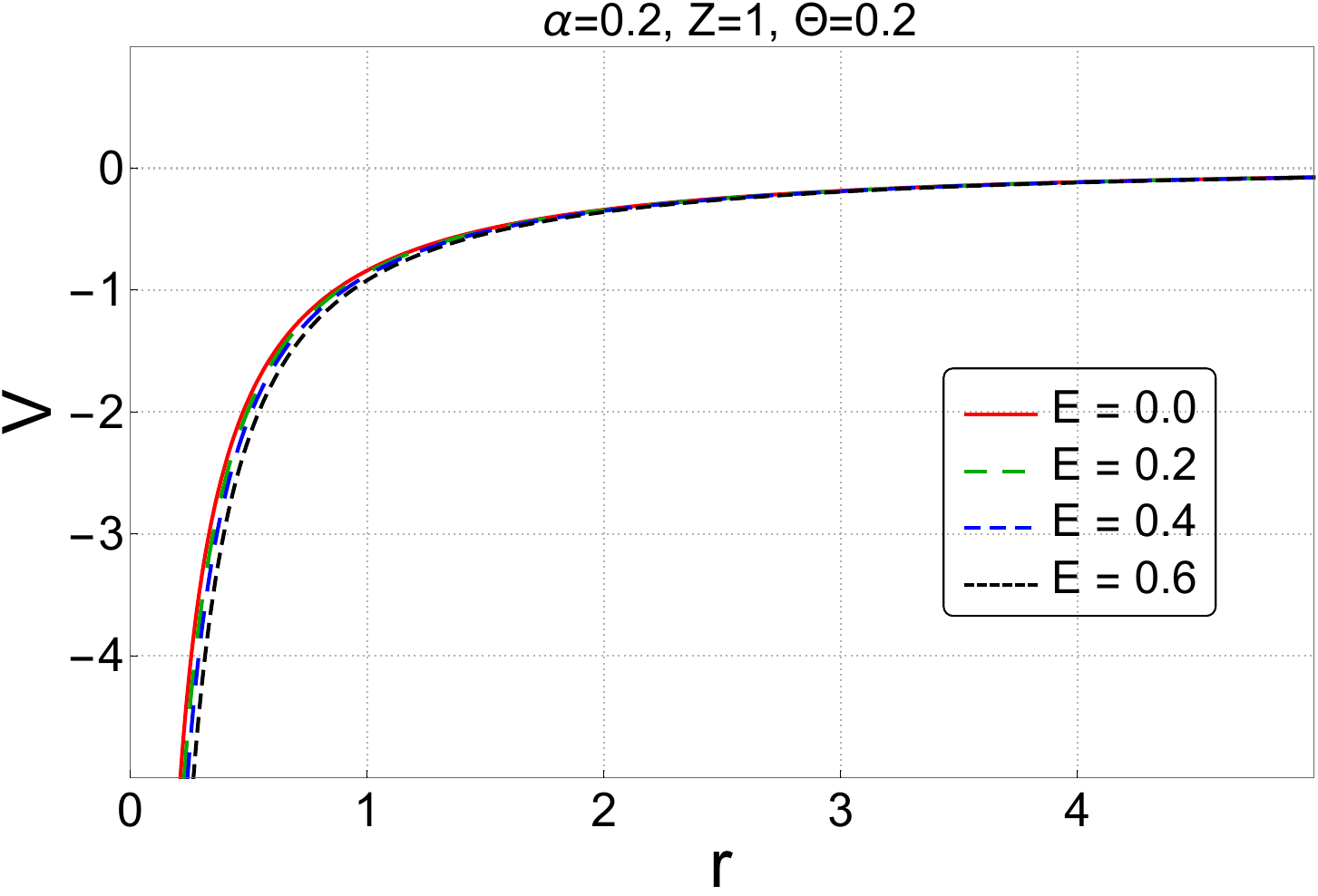}\hfill
	\includegraphics[width=0.25\textwidth]{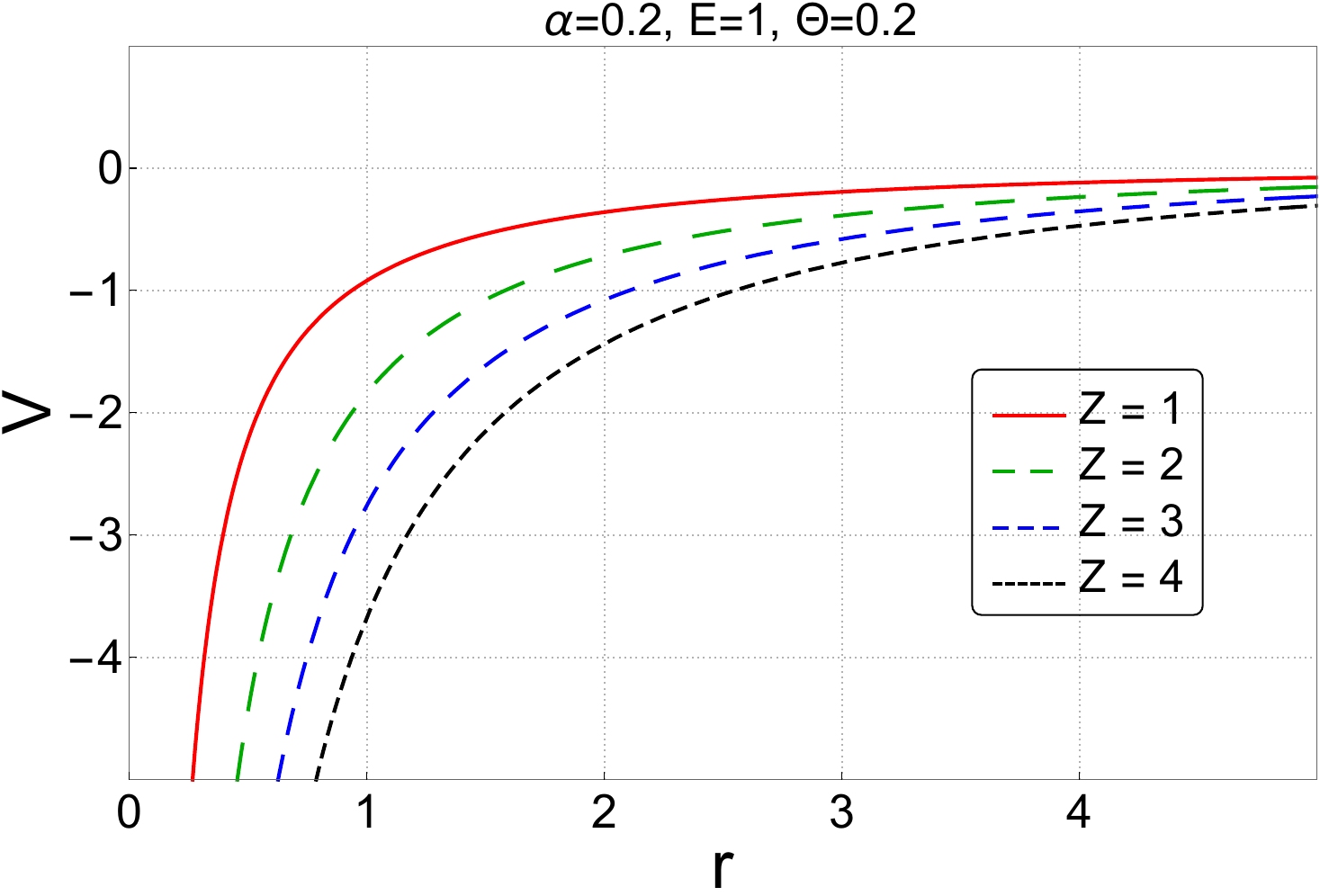}\hfill
	\includegraphics[width=0.25\textwidth]{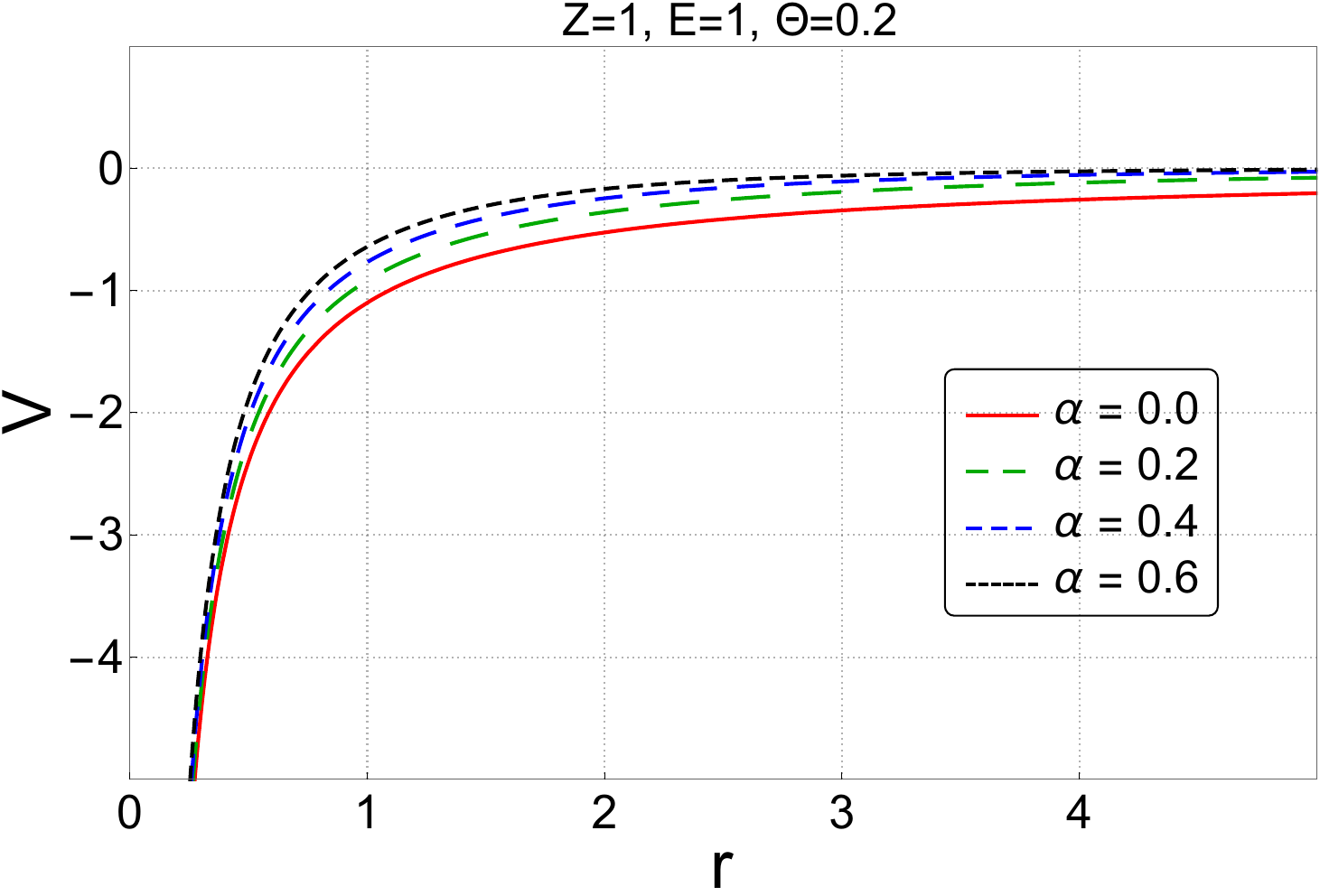}\hfill
	\includegraphics[width=0.25\textwidth]{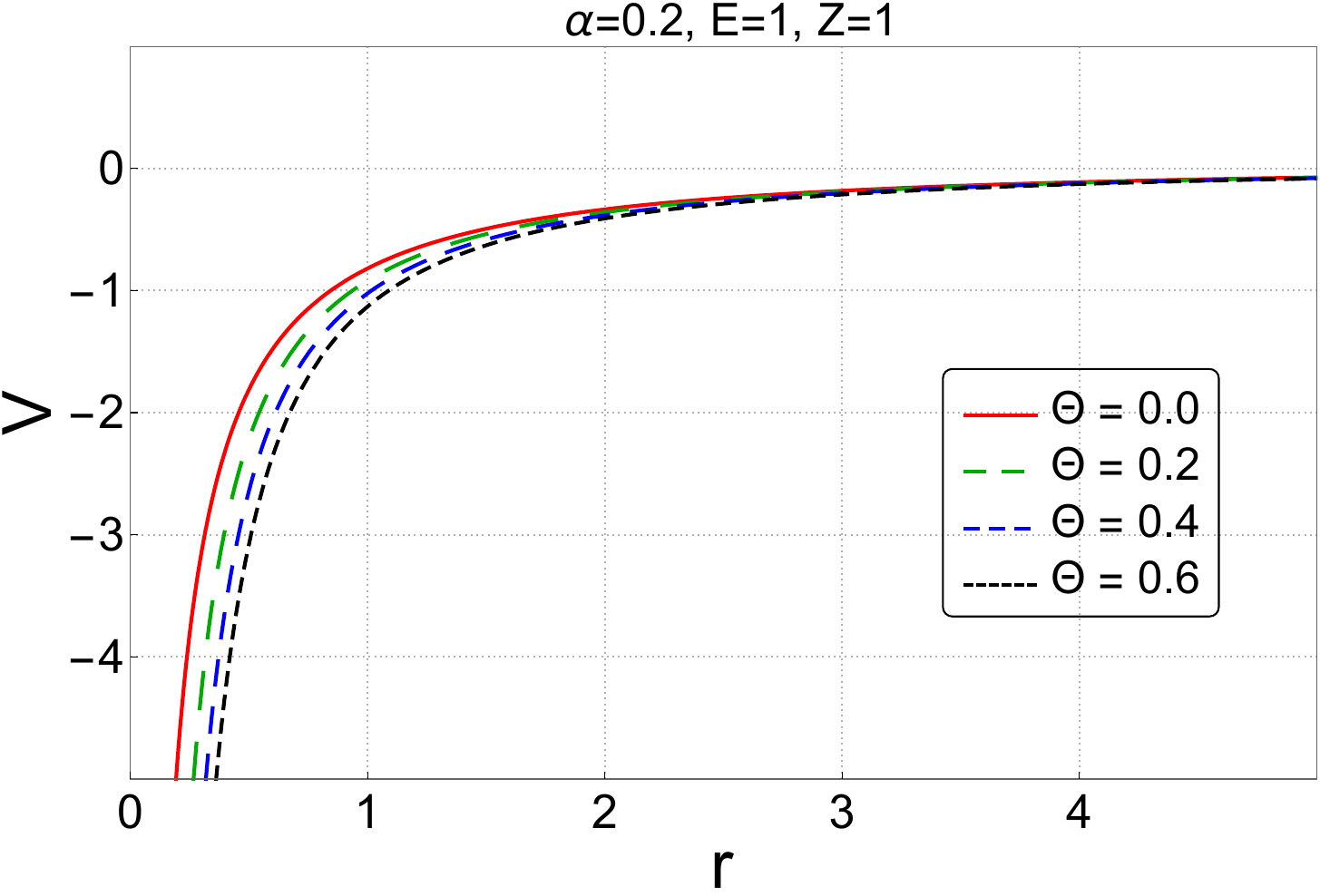}
	\caption{The behavior of the NC Yukawa potential as a function of $r$ and for different values of $E$, $Z$, $\alpha$ and $\Theta$. 
	}
	\label{fig1}
\end{figure*}
In Fig.~\ref{fig1}, we illustrate the behavior of the NC Yukawa potential as a function of $r$, $E$, $\alpha$, and $\Theta$ in different scenarios. From the figure, we observe that the energy of the particle and the NC parameter $\Theta$ influence the potential in a similar manner, resulting in an increase in the potential near the origin for increasing $E$ or $\Theta$. However, the effect of the screening parameter $\alpha$ differs, as an increase in $\alpha$ reduces the potential at larger distances. Overall, we find that the parameters $E,\,Z,\, \Theta$ tend to increase the potential, while $\alpha$ decreases it for most cases.

\begin{figure}[h]
	\centering
	\includegraphics[width=0.4\textwidth]{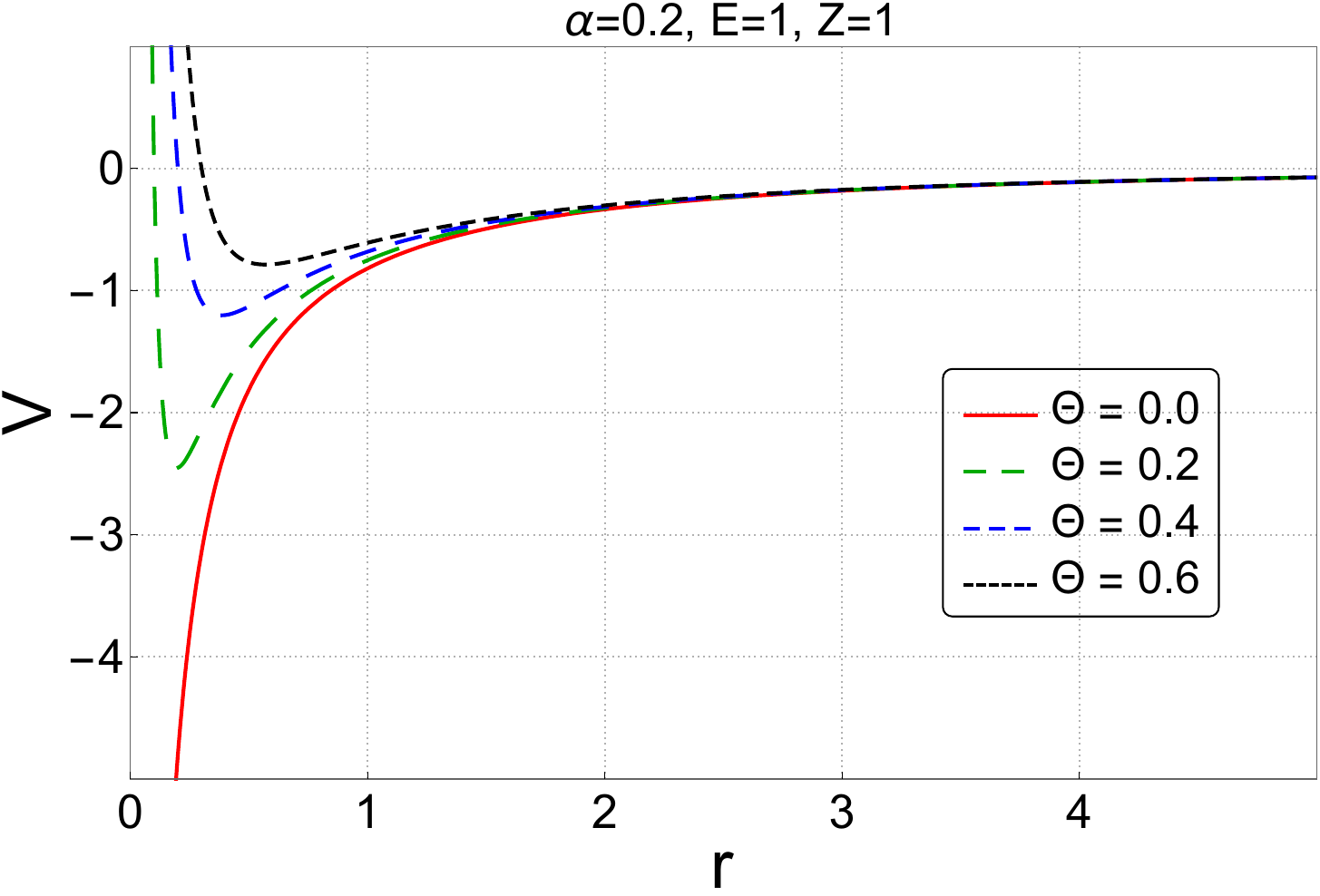}
	\caption{The behavior of the NC Yukawa potential as a function of $r$ and for different values of $\Theta$ in the case of the interaction with the atom electrons. }
	\label{fig1.2}
\end{figure}

The sign of $V_2$ can be positive ($V_0<0$) if we consider the potential in the same context as the Kratzer potential, which means that when accounting for the interaction of the incident electron with the atomic electrons, $V_0>0$, which is depicted in Fig. \ref{fig1.2}.

Additionally, the NC Yukawa potential can be reduced to the NC Coulomb potential \cite{moumni1}, which is structurally equivalent to the Kratzer potential when $\alpha=0$, and is given by \cite{kratzer1}:
\begin{equation}
	\hat{V}(r)=\left(\frac{V_0}{r}+\frac{V_1}{r^2}\right),\label{eq:YPNC4}
\end{equation}
where $V_1=\frac{V_0\Theta E}{2\hbar c}$ in this case. As expected, in the commutative limit $\Theta=0$, we recover the standard Coulomb potential. Thus, we see that the screened Kratzer and Kratzer potentials naturally emerge from the NC corrections to the Yukawa and Coulomb potentials, respectively.


\section{NC correction to the scattering amplitude $\hat{f}(\theta)$} \label{sec:NCCf}

In this section, we employ the Born approximation together with the NC potential \eqref{eq:YPNC2} to compute the NC scattering amplitude $\hat{f}(\theta)$ for electron scattering by the NC Yukawa potential. The scattering amplitude in the leading order of the Born approximation is expressed as \cite{cohen,cohen2}:
\begin{equation}
	\hat{f}(q)=-\frac{2m}{h^2}\frac{1}{q}\int_{0}^{\infty} \hat{V}(r) r \sin(qr) dr
\end{equation}
where $q = 2k \sin \frac{\theta}{2}$, and $k$ is the wave number of the incident electron. Using the modified potential form given in Eq.~\eqref{eq:YPNC3}, and performing the integration, we obtain:
\begin{equation}
	\hat{f}(q)=-\frac{2m V_1}{\hbar^2(q^2+\alpha^2)}-\frac{2mV_2}{\hbar^2q}\arctan\left(\frac{q}{\alpha}\right),\label{eq:NCf1}
\end{equation}
In terms of the NC parameter $\Theta$, the above expression can be rewritten as: 
	\begin{align}
		\hat{f}(\theta,\Theta)=-&\left(\frac{2m_eZ k_C e^2}{\hbar^2(4k^2\sin^2\frac{\theta}{2}+\alpha^2)}+\frac{m_eZk_C e^2 \Theta E}{2\hbar^2k\sin\frac{\theta}{2}}\right.\notag\\
		&\left.\times\arctan\left(\frac{2k\sin\frac{\theta}{2}}{\alpha}\right)\right)e^{\frac{1}{2\hbar c}\alpha \Theta E},\label{eq:NCf2}
	\end{align}
This expression represents the scattering amplitude for elastic scattering of electrons by the NC Yukawa potential, which is structurally similar to the screened Kratzer potential.

\subsection{Special cases} \label{subsec:SC}
We now examine particular cases of the NC amplitude for different potential types.

\subsubsection{Kratzer potential} \label{subsubsec:KP}
By setting $\alpha=0$, we obtain the scattering amplitude for the Kratzer potential:

\begin{equation}
	\hat{f}(\theta,\Theta)_{\text{Kratzer}}=-\frac{2m_e}{\hbar^2}\left(\frac{V_0}{4k^2\sin^2\frac{\theta}{2}}+\frac{V_1}{2k\sin\frac{\theta}{2}}\right),
\end{equation}
which is equivalent to the NC Coulomb potential with $V_0=Zk_C e^2$ and $V_1=V_0\Theta E/(2\hbar c)$.

\subsubsection{Yukawa potential} \label{subsubsec:YP}
By setting $V_2=0$ and $V_1=V_0$, we recover the scattering amplitude for the standard Yukawa potential:
\begin{equation}
	f(\theta)_{\text{Yukawa}}=-\frac{2m_e}{\hbar^2}\frac{V_0}{4k^2\sin^2\frac{\theta}{2}+\alpha^2},
\end{equation}
This corresponds to the elastic scattering amplitude for the Yukawa potential.

\subsubsection{Coulomb potential} \label{subsubsec:CP}

If we take $V_2=0$, $V_1=V_0$, and $\alpha=0$, we obtain the scattering amplitude for the Coulomb potential:
\begin{equation}
	f(\theta)_{\text{Coulomb}}=-\frac{2m_eV_0}{4\hbar^2k^2\sin^2\frac{\theta}{2}},
\end{equation}
The above equation represents the elastic scattering amplitude for the Coulomb potential.

These results demonstrate how the screened Kratzer potential reduces to the Coulomb potential under specific conditions, all originating from the NC geometry of spacetime. The NC corrections naturally yield the Kratzer potential form when the NC matrix $\Theta^{\mu\nu}$ is appropriately chosen as in Eq.~\eqref{eq:NCM}.


\section{Cross section of electron by atom in NC spacetime} \label{sec:NCCS}

To compute the cross section for electron scattering by an atom or a molecule, we employ the following expression for the differential cross section:
\begin{equation}
	\frac{d \hat{\sigma}}{d\Omega}=\left|\hat{f}(\theta,\Theta)\right|^2,
\end{equation}
where $d\Omega$ is the solid angle. By substituting the expression for the NC scattering amplitude from Eq.~\eqref{eq:NCf2}, the differential cross section in NC spacetime becomes:
\begin{widetext}
	\begin{equation}
		\frac{d \hat{\sigma}}{d\Omega}=\frac{4m_e^2}{\hbar^4}\left(\frac{V_1}{4k^2\sin^2\frac{\theta}{2}+\alpha^2}+\frac{V_2}{2k\sin\frac{\theta}{2}}\arctan\left(\frac{2k\sin\frac{\theta}{2}}{\alpha}\right)\right)^2,
	\end{equation}
\end{widetext}
which corresponds to the scattering of an electron by the NC Yukawa potential, where the parameters are defined as $V_1=V_0 e^{\frac{\alpha \Theta E}{2\hbar c}}$ and $V_2=V_0 e^{\frac{\alpha \Theta E}{2\hbar c}} \frac{\Theta E}{2\hbar c}$. From the expression above, we can recover the previously discussed special cases: For $V_2=0$ and $V_1=V_0$, we obtain the standard Yukawa scattering. Setting $\alpha=0$ reduces the result to the Coulomb scattering. For the case $V_2=\pi V_0\Theta E/(4\hbar c)$ and $V_1=V_0$, we obtain the Kratzer scattering amplitude:
\begin{equation}
	\frac{d \hat{\sigma}}{d\Omega}=\frac{4m_e^2}{\hbar^4}\left(\frac{V_1}{4k^2\sin^2\frac{\theta}{2}+\alpha^2}+\frac{V_2}{2k\sin\frac{\theta}{2}}\right)^2,
\end{equation}
	\begin{figure*}[]
		\centering
		\includegraphics[width=0.23\textwidth]{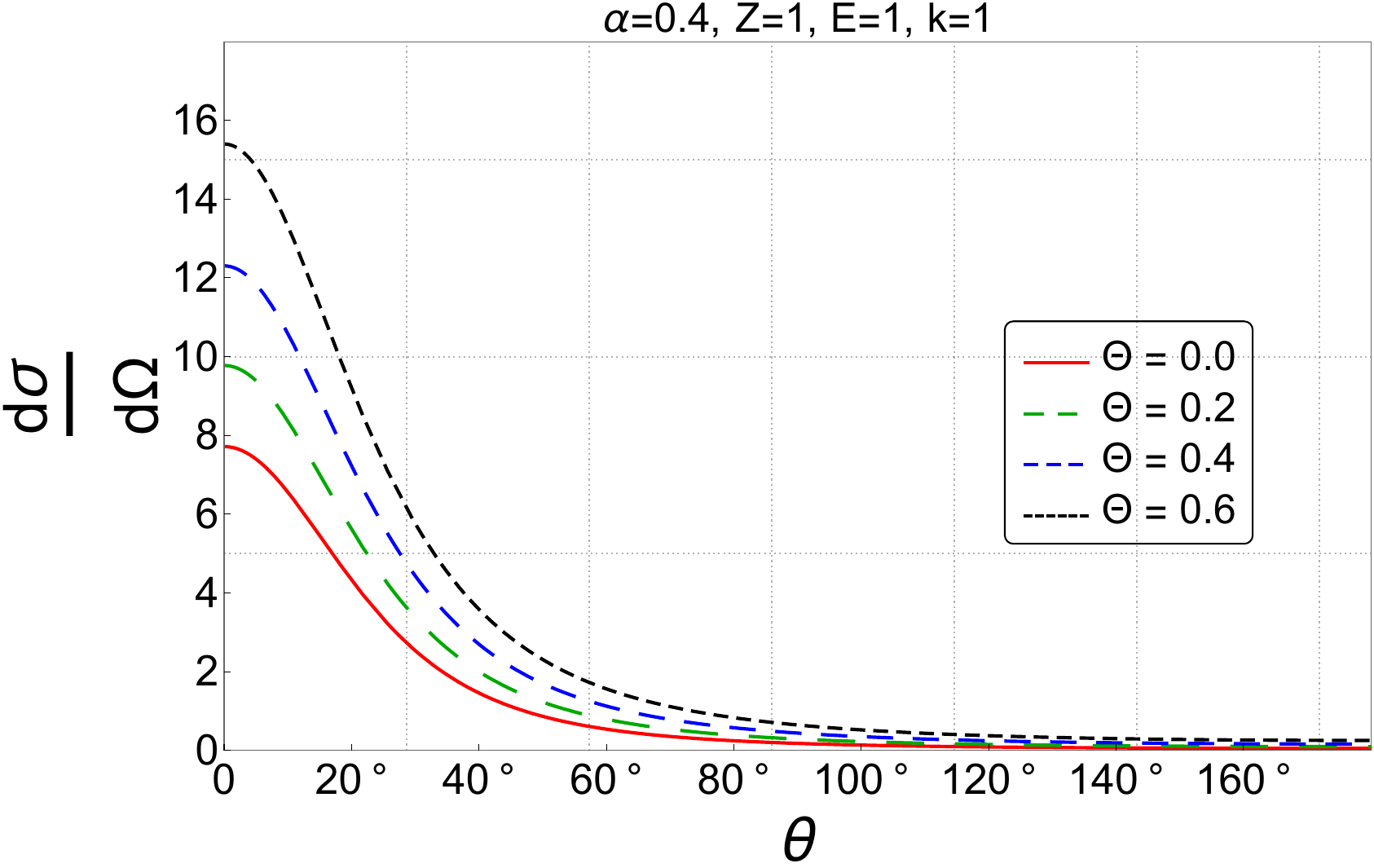}\hfill
		\includegraphics[width=0.23\textwidth]{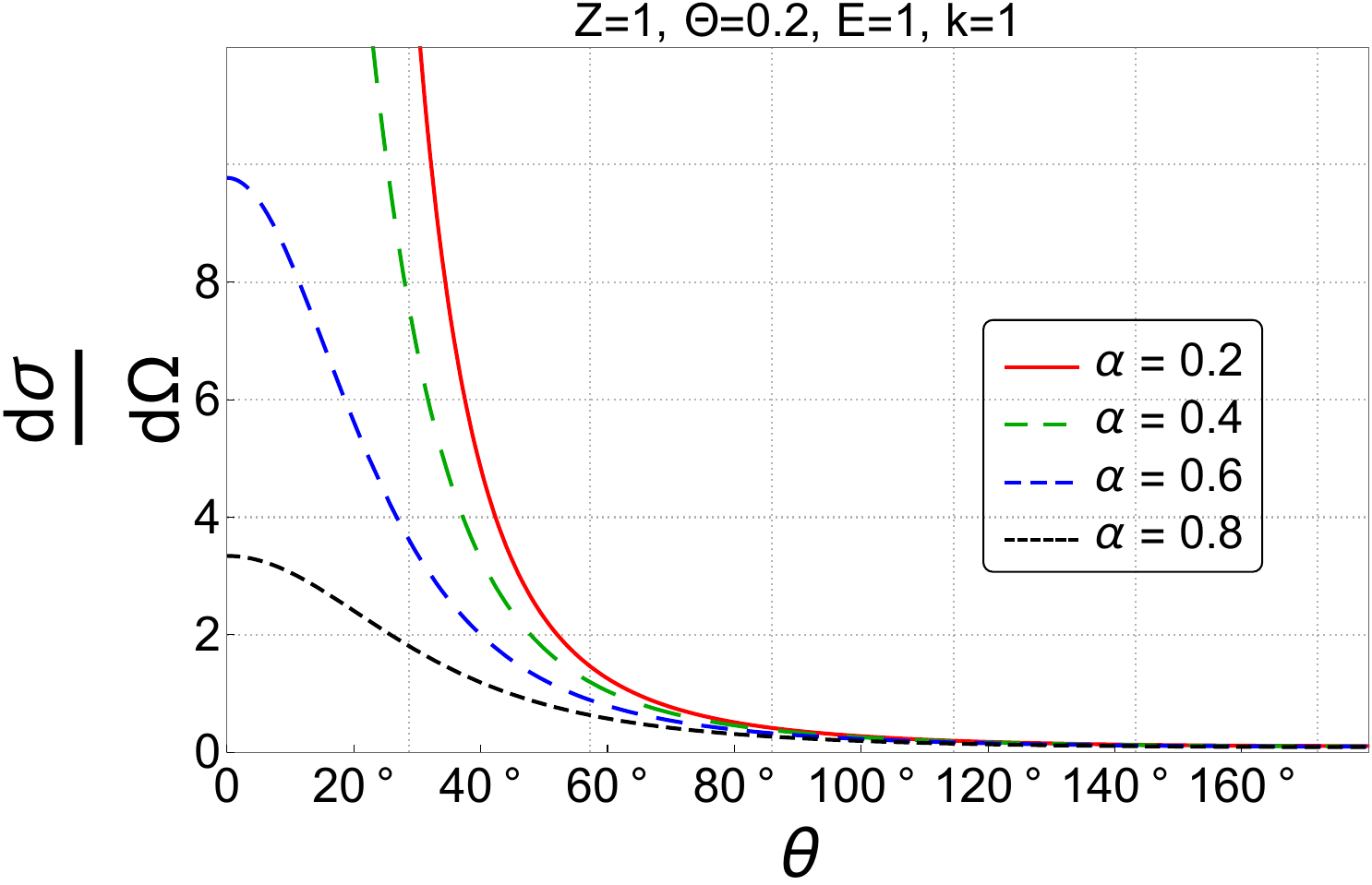}\hfill
		\includegraphics[width=0.23\textwidth]{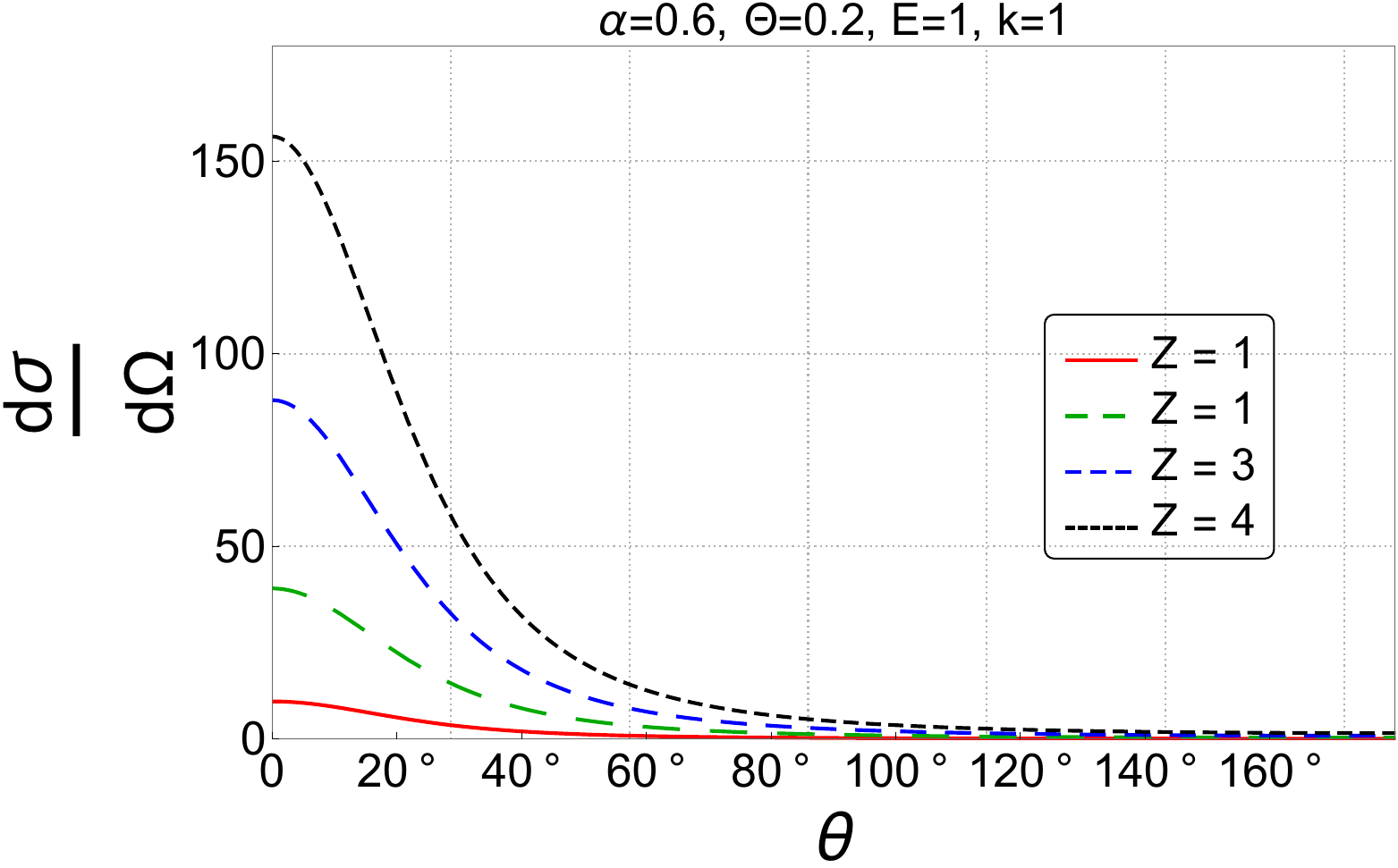}\hfill
		\includegraphics[width=0.23\textwidth]{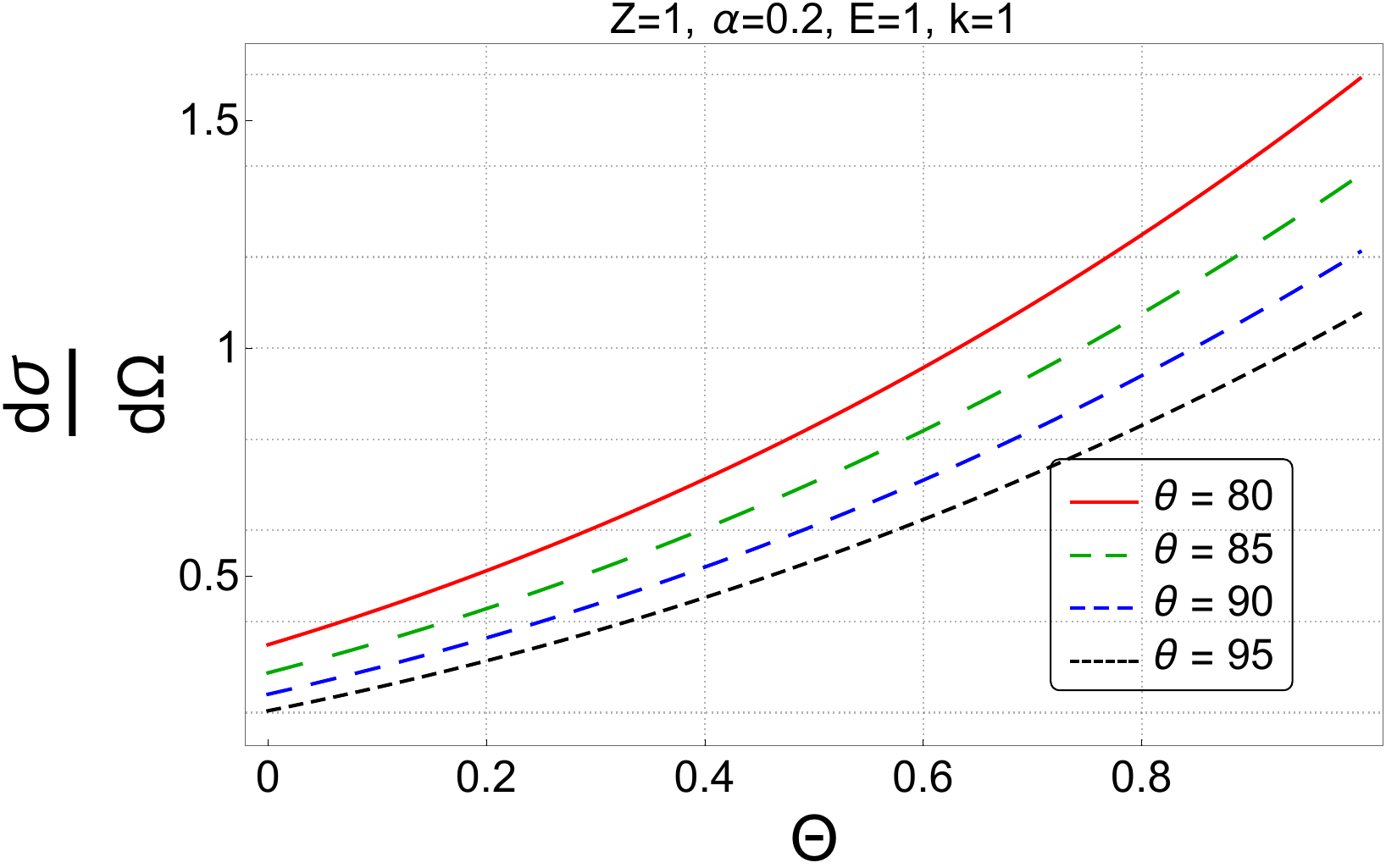}\hfill
		\caption{The behavior of the NC differential cross section as a function of $\theta$ and for different values of  $\Theta$, $\alpha$ and $Z$ (first three panel). As a function of $\Theta$ and for different value of the angle $\theta$.}
		\label{fig2}
	\end{figure*}

To obtain the total cross section using the first-order Born approximation, we use the following integral representation:
\begin{equation}
	 \hat{\sigma}=\frac{2\pi}{k^2}\int_{0}^{2k}\left|\hat{f}(q)\right|^2\, q\,dq,
\end{equation}
By substituting the scattering amplitude from Eq.~\eqref{eq:NCf1}, we obtain:
\begin{widetext}
\begin{align}
	\hat{\sigma}=&\frac{8\pi m_e^2}{\hbar^4}\left(\frac{2 V_1^2}{\alpha^2(4k^2+\alpha^2)}+\frac{V_1V_2}{15\,\alpha^5 k^2}\left(\alpha^4\log\left(\frac{4k^2}{\alpha^2}+1\right)+24k^4-32k^2\alpha^2\right)\right.\notag\\
	&\left.+\frac{2V_2^2}{135\,\alpha^6}\left(368k^4-180k^2\alpha^2+135\alpha^4\right)\right),
\end{align}
\end{widetext}
The above expression represents the total cross section for elastic scattering of an electron by the Yukawa potential in NC spacetime\footnote{We have using the following approximation $\arctan\left(\frac{q}{\alpha}\right)\approx \frac{q}{\alpha} - \frac{q^3}{3\alpha^3}+\frac{q^5}{\alpha^3}+ \mathcal{O}\left(\left(\frac{q}{\alpha}\right)^7\right)$.}. If we consider the interaction of the incident electron with the electrons in the atom, the sign of $V_0$ should be taken as positive, $V_0 > 0$, in the component $V_2$.

\begin{figure*}[]
	\centering
	\includegraphics[width=0.33\textwidth]{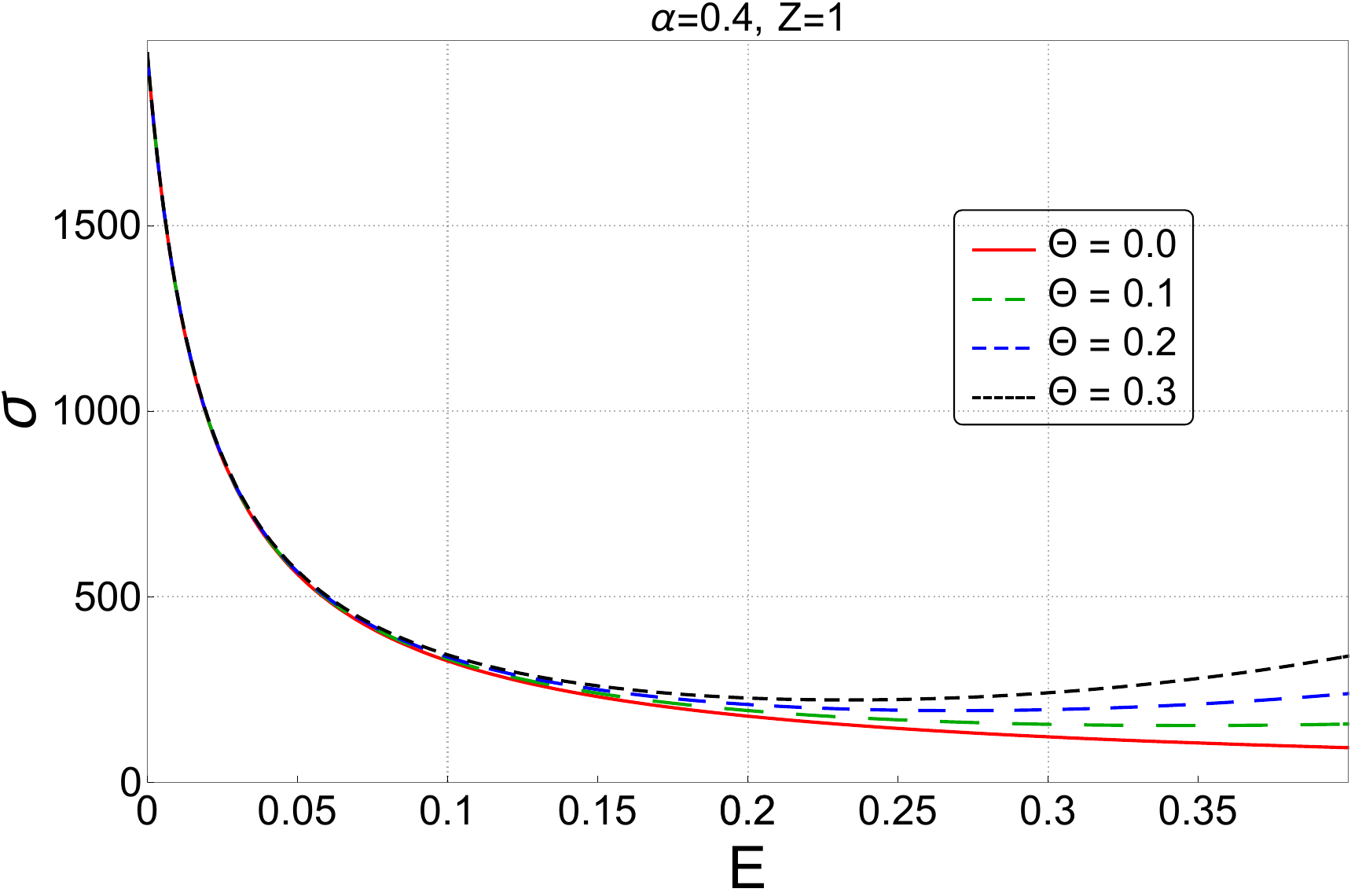}\hfill
	\includegraphics[width=0.33\textwidth]{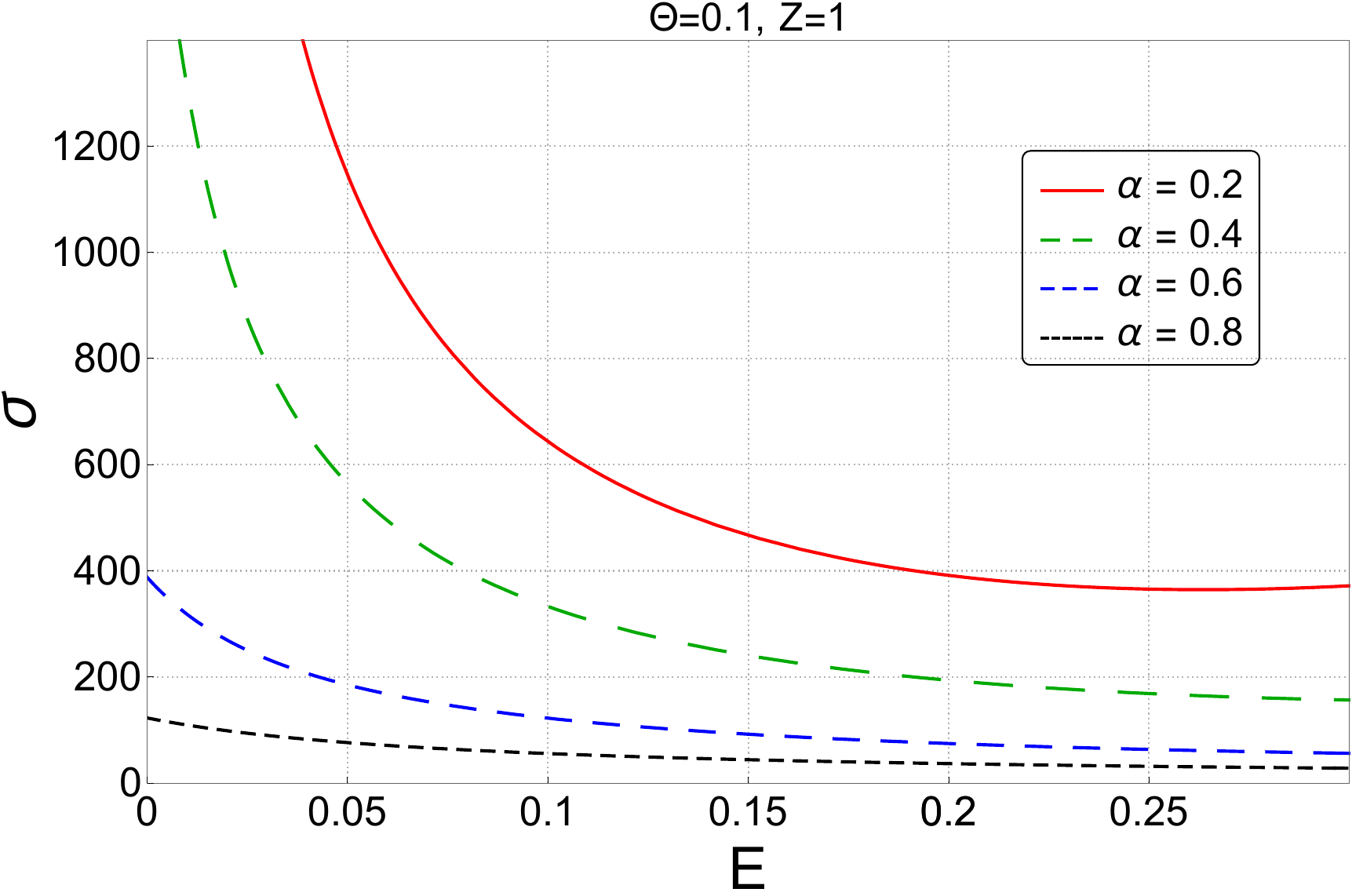}\hfill
	\includegraphics[width=0.33\textwidth]{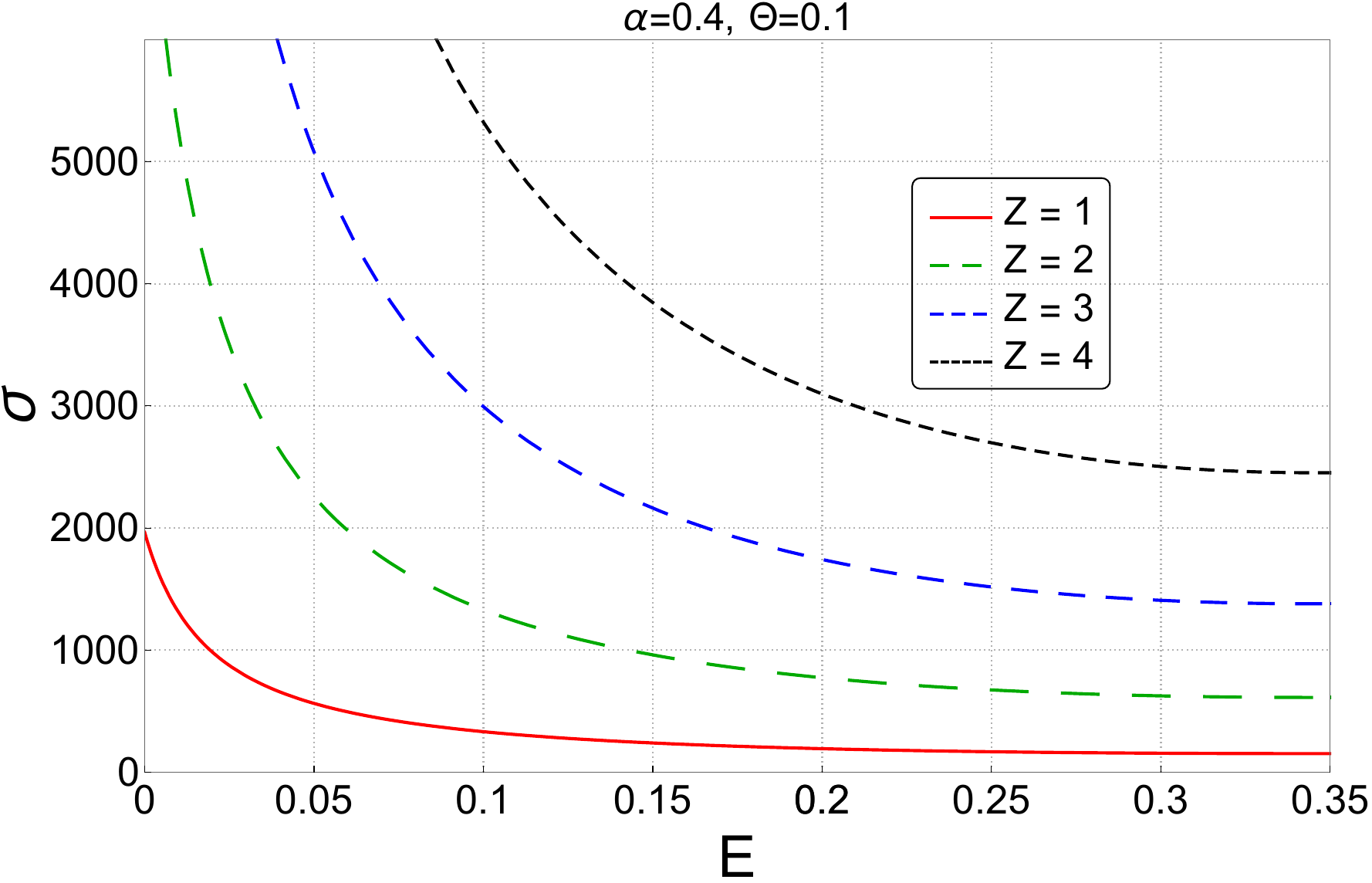}
	\caption{The behavior of the NC total cross section as a function of $E$ and for different values of  $\Theta$, $\alpha$ and $Z$.}
	\label{fig3}
\end{figure*}

			
		
		
			
			


\subsection{Application to Scattering by Molecules} \label{subsec:Application}

In this subsection, we discuss the constraints on the NC parameter $\Theta$ by analyzing the elastic scattering of electrons by molecules, focusing on various regimes of incident electron energy. Specifically, we examine how variations in energy affect the bounds on the NC parameter, offering insights into its behavior under different experimental conditions.

\begin{figure*}[]
	\centering
	\includegraphics[width=0.23\textwidth]{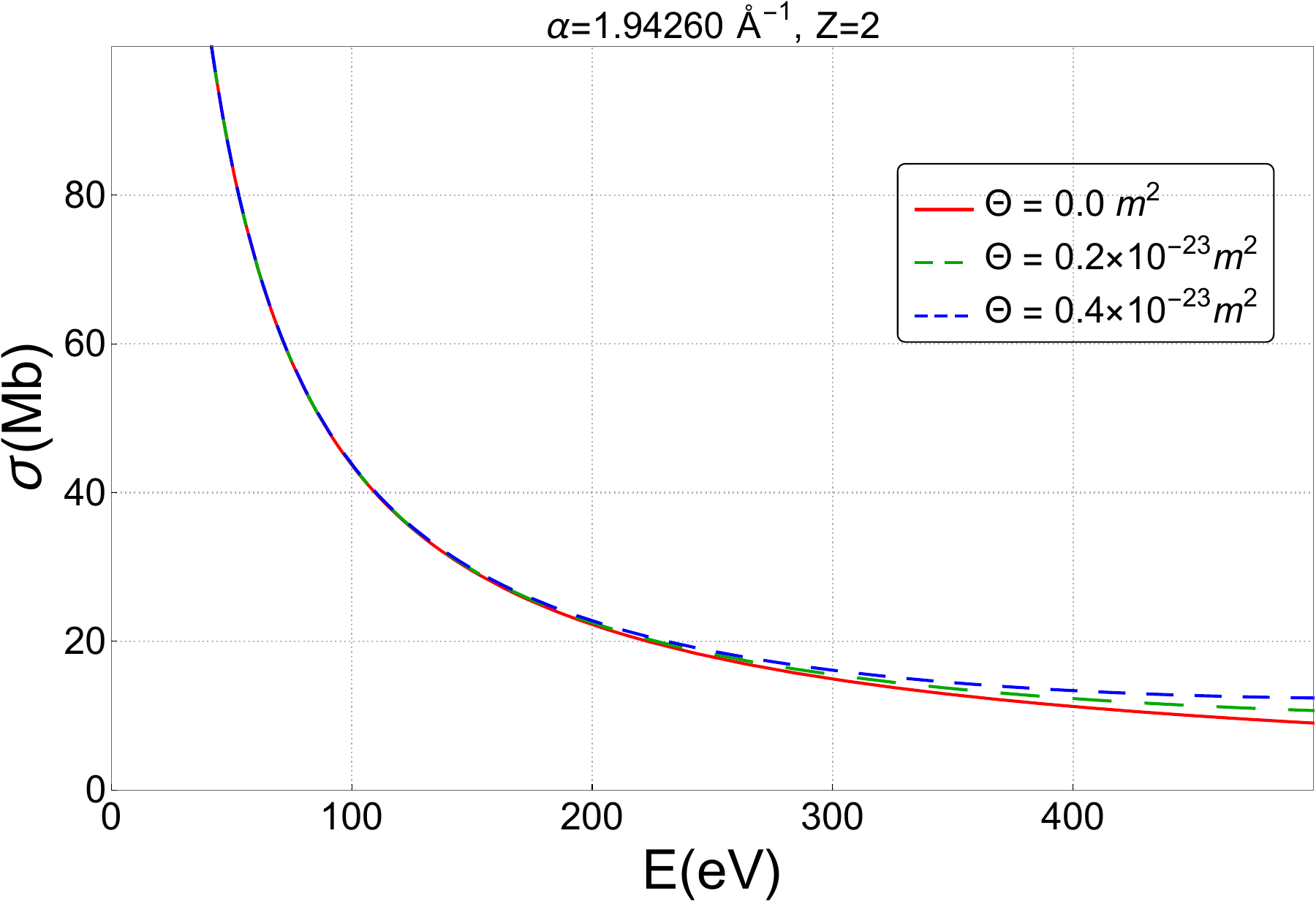}
	\includegraphics[width=0.23\textwidth]{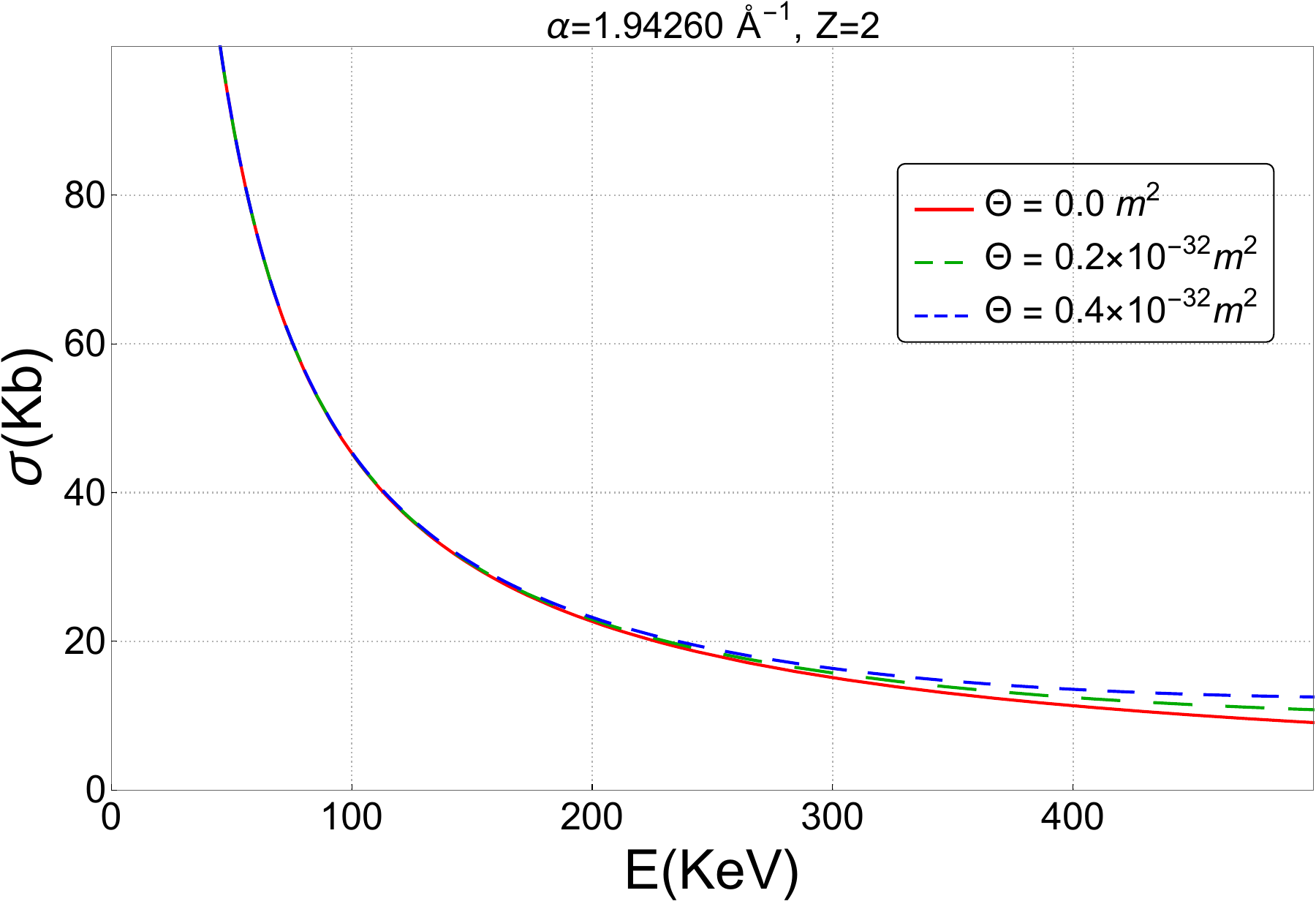}
	\includegraphics[width=0.23\textwidth]{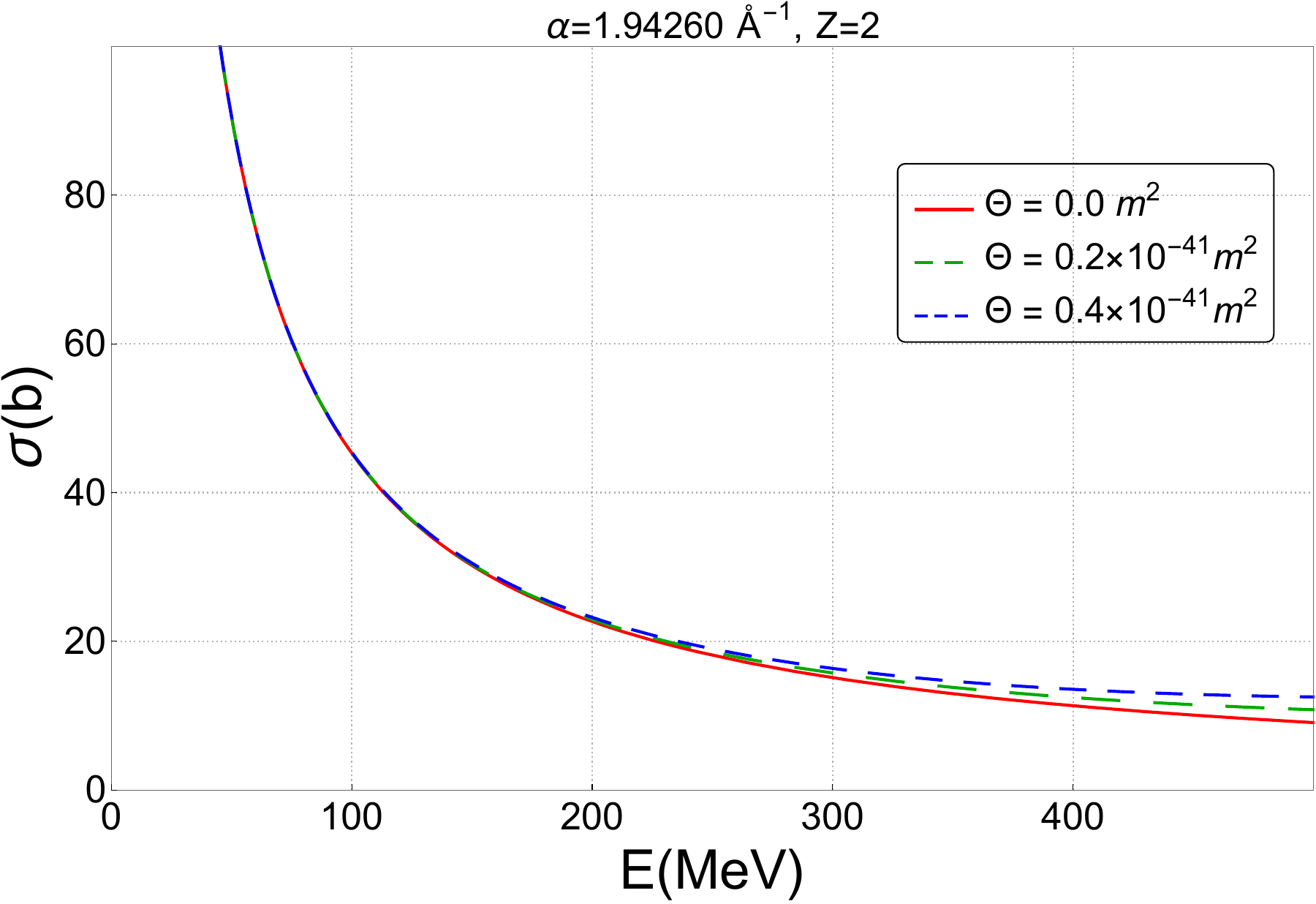}
	\includegraphics[width=0.23\textwidth]{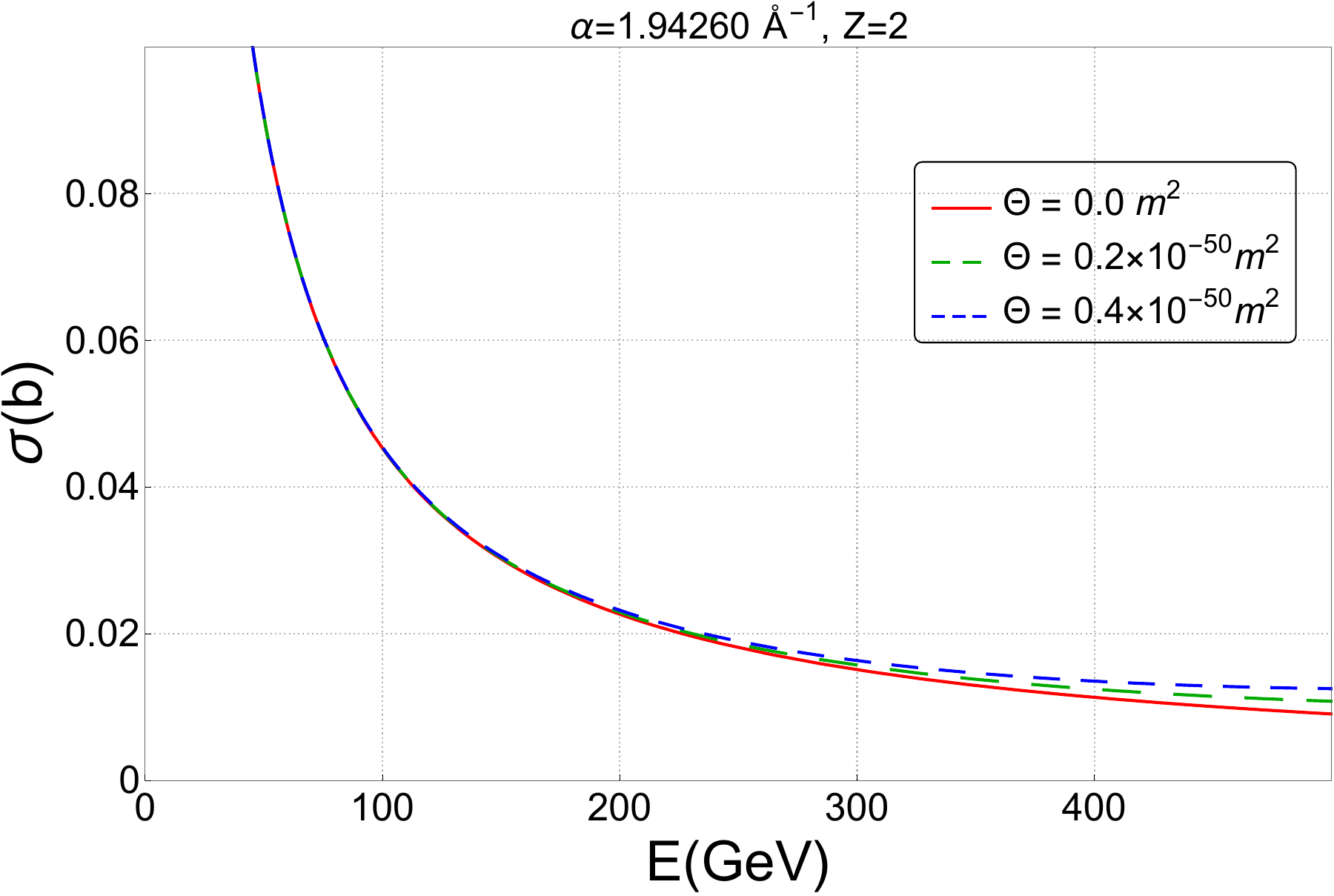}
	\caption{The behavior of the NC total cross section as a function of energy $E$ for different values of the NC parameter $\Theta$, in the case of electron scattering by an $H_2$ molecule, from low to high energy levels (left to right panels).}
	\label{fig4}
\end{figure*}

Figure \ref{fig4} shows the behavior of the total cross section as a function of the electron's energy when scattered by an $H_2$ molecule, for various values of the NC parameter $\Theta$. The figure covers a wide range of electron energy levels, from non-relativistic to ultra-relativistic regimes (from eV to GeV). The influence of non-commutativity appears consistently across all energy levels. Notably, in the non-relativistic regime, the NC effect becomes significant for relatively large values of $\Theta$, whereas in the ultra-relativistic regime, the NC effects are prominent only for much smaller values of $\Theta$ compared to the non-relativistic case. This reveals how energy impacts the description of the fundamental properties of spacetime.

\begin{table}[]
	\centering
	\begin{tabular}{ c | c | c  }
		\hline	\hline
		Target	& Electron Energy & Bound on $\Theta$ 
		\\
		\hline	
		\hline
		& 1 eV & $\sqrt{\Theta} \geq 10^{-11}\,\text{m}$  \\
		
		& 1 KeV  &  $\sqrt{\Theta} \geq 10^{-15}\,\text{m}$   \\
		
		$H_2$	& 1 MeV & $\sqrt{\Theta} \geq 10^{-19}\,\text{m}$  \\
		
		& 1 GeV & $\sqrt{\Theta} \geq 10^{-24}\,\text{m}$\\
		
		& 100 GeV &  $\sqrt{\Theta} \geq 10^{-26}\,\text{m}$ \\
		
		\hline
		\hline
	\end{tabular}
	\caption{Estimates of the NC parameter for different electron energy levels in elastic scattering by an $H_2$ molecule with $Z=2$ and screened parameter $\alpha = 1.9426\,\AA^{-1}$ \cite{oluwadare1}.}	
	\label{tableau1}
\end{table}

Table. \ref{tableau1} presents the lower bounds on $\Theta$ obtained from electron scattering by an $H_2$ molecule for various energy levels. The spectroscopic parameters of the target molecule allow us to estimate these bounds for different incident electron energies. As shown, the NC parameter from the scattering of non-relativistic electrons is on the order of $10^{-11}\,\text{m}$. For ultra-relativistic electrons, the lower bound on $\sqrt{\Theta}$ reaches the order of $10^{-26}\,\text{m}$. Similar bounds are obtained for light molecules, while for heavier molecules, such as Scandium hydride ($ScH$, with atomic number $Z=22$ and screened parameter $\alpha = 1.41113\,\AA^{-1}$ \cite{onyeaju1}), the lower bound on $\sqrt{\Theta}$ decreases to $10^{-28}\,\text{m}$ for ultra-relativistic incident electrons. 

It is noteworthy that heavier molecules consistently provide lower bounds on $\Theta$, with the smallest values observed at energy levels around 100 GeV. These results provide a robust lower bound on the NC parameter in quantum systems, significantly improving upon the bounds obtained from NC field theory \cite{NCfield1,NCfield2}, which range from $10^{-17}\,\text{m}$ to $10^{-20}\,\text{m}$. Additionally, bounds on the NC parameter have been studied using Hydrogen atoms and the Lamb shift \cite{NCHlamb1,NCHlamb2,zaim9,zaim8,zaim7,zaim2,zaim4,zaim3,zaim1}, where $\Theta$ is estimated to be around $10^{-15}\,\text{m}$ to $10^{-22}\,\text{m}$. Furthermore, studies of particle decay and collision processes using the NC Standard Model \cite{NCdecay1,NCcoll1,NCcoll2,NCcoll3,NCcoll4} place bounds on $\Theta$ between $10^{-16}\,\text{m}$ and $10^{-19}\,\text{m}$. Our results, based on elastic electron scattering by heavy molecules, show a new lower bound on $\Theta$ at the level of $10^{-28}\,\text{m}$, representing a significant improvement in quantum systems.

Remarkably, the bounds obtained from macroscopic systems, such as classical tests of general relativity \cite{abdellah1,abdellah2}, black hole thermodynamics \cite{abdellah3,abdellah4,abdellah5}, and gravitational wave studies \cite{NCGW1}, offer even tighter constraints on $\Theta$, ranging from $10^{-31}\,\text{m}$ to $10^{-35}\,\text{m}$. These tighter bounds are due to the higher energy levels involved in these systems, which depend on the energy scale of the phenomena under consideration.


\section{Conclusions}\label{sec:concl}

In this work, we have performed a detailed investigation of the differential and total cross sections for the elastic scattering of electrons by a NC Yukawa potential.

First, we derived the NC correction to the Yukawa potential up to first order in the NC parameter. Our findings suggest that the modified Yukawa potential resembles a screened Kratzer potential, which naturally emerges from the NC nature of spacetime. We demonstrated that, in the appropriate limit, the NC extension of the Yukawa potential reduces to the well-known NC Coulomb potential. This reduction leads to the Kratzer potential form, revealing that both the Kratzer and screened Kratzer potentials can be interpreted as geometric consequences of spacetime non-commutativity. As a result, this new potential unifies four distinct potentials—screened Kratzer, Kratzer, Yukawa, and Coulomb—under a single mathematical framework in three limiting cases.

Second, we employed the first-order Born approximation to compute the scattering amplitude for the NC Yukawa potential. Through this approach, we derived the scattering amplitudes for the Kratzer, Yukawa, and Coulomb potentials as specific cases.

Finally, we calculated the differential and total cross sections for the elastic scattering of electrons by this new NC potential. As a practical application, we analyzed the scattering of electrons by an $H_2$ molecule to estimate the NC parameter based on experimental data at different incident electron energy levels. Our results indicate a direct relationship between the NC parameter and the energy of the system. This is reminiscent of the behavior of gravity at macroscopic scales, where the curvature of spacetime is related to the mass of an object. At the microscopic level, non-commutativity similarly exhibits a direct relationship between spacetime and the system's energy. The higher the energy, the closer the NC parameter approaches the Planck scale. For non-relativistic electrons, we obtained a bound of the order of $10^{-11}\,\text{m}$, while for ultra-relativistic electrons, we derived a bound of $10^{-26}\,\text{m}$ (see Table \ref{tableau1}). Additionally, for a heavy molecule like $ScH$, we achieved a new bound of magnitude $10^{-28}\,\text{m}$ for ultra-relativistic incident electrons.

Furthermore, the non-commutativity of spacetime can be interpreted as a form of "tension" in spacetime, whereby spacetime responds to the presence of energy at the quantum level, similar to its response to matter at the macroscopic level. This highlights the potential of NC geometry as a viable model for describing quantum gravity, providing a pathway to quantizing gravity at quantum scales.


\bibliography{ref}

\end{document}